\documentclass[trackchanges,twocolumn]{aastex701}
\usepackage{mathrsfs}

\def\hal{H$\alpha$}
\def\hb{H$\beta$}

\def\oiii{[\ion{O}{3}]}
\def\nii{[\ion{N}{2}]}
\def\feii{\ion{Fe}{2}}
\def\sii{[\ion{S}{2}]}

\begin{document}

\title{Probing the Physical Origin of the Balmer Decrement in the Broad-line Region of Nearby Active Galactic Nuclei via Spectral Variability }

\author[0000-0002-5346-0567]{Suyeon Son}
\affiliation{Kavli Institute for Astronomy and Astrophysics, 
Peking University, Beijing 100871, People's Republic of China}
\affiliation{Department of Astronomy and Atmospheric Sciences, 
Kyungpook National University, Daegu 41566, Republic of Korea}
\email[show]{sson.astro@gmail.com}

\author[0000-0002-3560-0781]{Minjin Kim}
\affiliation{Department of Astronomy, Yonsei University, 
50 Yonsei-ro, Seodaemun-gu, Seoul 03722, Republic of Korea}
\affiliation{Department of Astronomy and Atmospheric Sciences, 
Kyungpook National University, Daegu 41566, Republic of Korea}
\email[show]{mkim.astro@yonsei.ac.kr}

\author[0000-0001-6947-5846]{Luis C. Ho}
\affiliation{Kavli Institute for Astronomy and Astrophysics, Peking University, Beijing 100871, People's Republic of China}
\affiliation{Department of Astronomy, School of Physics, Peking University, Beijing 100871, People's Republic of China}
\email{lho.pku@gmail.com}

\author[0000-0001-8496-4162]{Ruancun Li}
\affiliation{Kavli Institute for Astronomy and Astrophysics, Peking University, Beijing 100871, People's Republic of China}
\affiliation{Department of Astronomy, School of Physics, Peking University, Beijing 100871, People's Republic of China}
\email{liruancun@gmail.com }

\begin{abstract}

To investigate the physical origin of the Balmer decrement in the broad-line region of active galactic nuclei (AGNs), we measure the temporal variability of the fluxes of the broad \hb\ and \hal\ emission lines using multi-epoch spectroscopic data of low-redshift AGNs from the Sloan Digital Sky Survey. The analysis of the mean spectra reveals that the Balmer decrement shows no correlation with AGN luminosity, while it is inversely correlated with the Eddington ratio. However, the temporal variation of the Balmer decrement in individual objects exhibits an even stronger anti-correlation with AGN luminosity, suggesting that the change in AGN luminosity plays a dominant role in determining the Balmer decrement. By comparing the temporal evolution of the Balmer decrement with the continuum color, we find that reddening due to the AGN itself may not be the primary factor. Instead, radiative transfer effects and excitation mechanisms, which deviate from the Case B recombination, appear to be critical for the variation of the Balmer decrement. These results provide useful insights into the underlying physics of changing-look AGNs and high-$z$ AGNs, such as the ``little red dots'', which exhibit extreme values of the Balmer decrement that can be misinterpreted as evidence for dust.
\end{abstract}


\section{Introduction} 
The hydrogen recombination lines are among the strongest features in both the broad-line region (BLR) and narrow-line region (NLR) of active galactic nuclei (AGNs). Their relative strengths, namely the Balmer decrement, have conventionally been used as indicators of dust extinction by comparing them with predictions from theoretical models. The Balmer decrement is widely employed because of its relative insensitivity to gas conditions such as temperature and density, and the \hal/\hb\ ratio is most often used because both lines are relatively strong and commonly accessible. For photoionized nebulae, the intrinsic Balmer decrement is calculated under the assumption of Case B recombination, in which the gas density is high enough to be optically thick to Lyman emission \citep{baker_1938}. The intrinsic NLR Balmer decrement is typically $\sim3.0-3.1$, marginally higher than the Case B prediction due to enhancement of \hal\ by collisional excitation \citep{halpern_1982,halpern_1983,gaskell_1984,osterbrock_2006}. Since the intrinsic Balmer decrement exhibits little scatter in low-density environments such as the NLR, planetary nebulae, and \ion{H}{2} regions, an observed steep Balmer decrement is thought to be primarily caused by dust reddening.

In contrast, the intrinsic Balmer decrement in the BLR can be further influenced by gas conditions, such as temperature and density, due to its high density \citep{osterbrock_2006}. A broad range of BLR Balmer decrements, ${\rm H\alpha}/{\rm H\beta} \approx 2.7$ to $3.5$, has been reported for different type 1 AGN samples \cite[e.g.,][]{greene_2005, lamura_2007,dong_2008,gaskell_2017,lu_2019b,sriram_2022}.
Photoionization models have successfully reproduced the observed range of Balmer decrements without invoking reddening, as the Balmer decrement is highly sensitive to gas conditions and the strength and shape of the ionizing continuum in high-density environments such as the BLR \cite[e.g.,][]{netzer_1975,davidson_1979,wu_2023}. 
Consequently, the mechanism behind the variation of the Balmer decrement in the BLR, whether due to reddening or other processes, remains controversial.

There have been attempts to investigate the correlation between the Balmer decrement and the accretion state of the AGN, using observed power-law indices or continuum slopes, ranging from the optical to the X-rays \citep{dong_2008,baron_2016,sriram_2022}. \cite{baron_2016} and \cite{sriram_2022} found that the Balmer decrement increases when the optical continuum slope ($L_{\nu} \propto \nu^\alpha_{\rm opt}$ between 3000 and 5100 \AA) becomes redder and the X-ray photon index at $2-10$ keV becomes harder. They attributed this correlation to dust obscuration, which becomes stronger when the optical continuum reddens from dust and the X-ray photon index hardens from photoelectric absorption. This interpretation, however, conflicts with other studies that failed to find a clear connection between the Balmer decrement and luminosity \citep{dong_2008,lu_2019b}, although the torus covering factor is known to increase with decreasing luminosity (e.g., \citealt{maiolino_2007,mor_2011,lusso_2013,trefoloni_2025}; but see \citealt{netzer_2016,stalevski_2016}). Instead, the Balmer decrement has been reported to be larger for AGNs with lower Eddington ratios  \cite[e.g.,][]{lamura_2007,sriram_2022,lu_2019b}, consistent with the observed trend that the torus covering factor becomes larger with decreasing Eddington ratio for AGNs with intermediate Eddington ratios \cite[e.g.,][]{ricci_2017c,zhuang_2018,ricci_2023}. On the other hand, \cite{wu_2023} argued that the inverse correlation
between Balmer decrement and Eddington ratio is unlikely to be driven by dust obscuration, as low-Eddington ratio AGNs with minimal absorption still exhibit steep Balmer decrements. Furthermore, the presence of intermediate-type AGNs (type 1.8/1.9; \citealt{osterbrock_1981}), characterized by elevated Balmer decrements despite weak X-ray absorption, indicates dust obscuration cannot fully account for the steep Balmer decrement \cite[e.g.,][]{barcons_2003}.

Disentangling the contributions of accretion state and dust obscuration from mean Balmer decrement measurements is not straightforward. Temporal variations of the Balmer decrement offer an alternative diagnostic, although previous studies have mostly focused on individual sources, including changing-look AGNs. The Balmer decrement of AGNs is observed to rise when the continuum or \hb\ flux drops \cite[e.g.,][]{antonucci_1983,shapovalova_2004,kollatschny_2018,kollatschny_2022,li_2022}, but its behavior can be more complex depending on the AGN state. For example, in their 20-year spectral variability study of the Seyfert 1 galaxy NGC 7603, \cite{kollatschny_2000} discovered that the BLR Balmer decrement increases as the \hb\ intensity decreases in the low state, but this trend disappears in the high state, leaving the physical origin of these empirical trends unresolved. Similar state-dependent behavior of the Balmer decrement in response to continuum flux has also been reported in other objects \cite[e.g.,][]{shapovalova_2010,sergeev_2011,popovic_2011}.

\begin{figure*}[tp]
\plotone{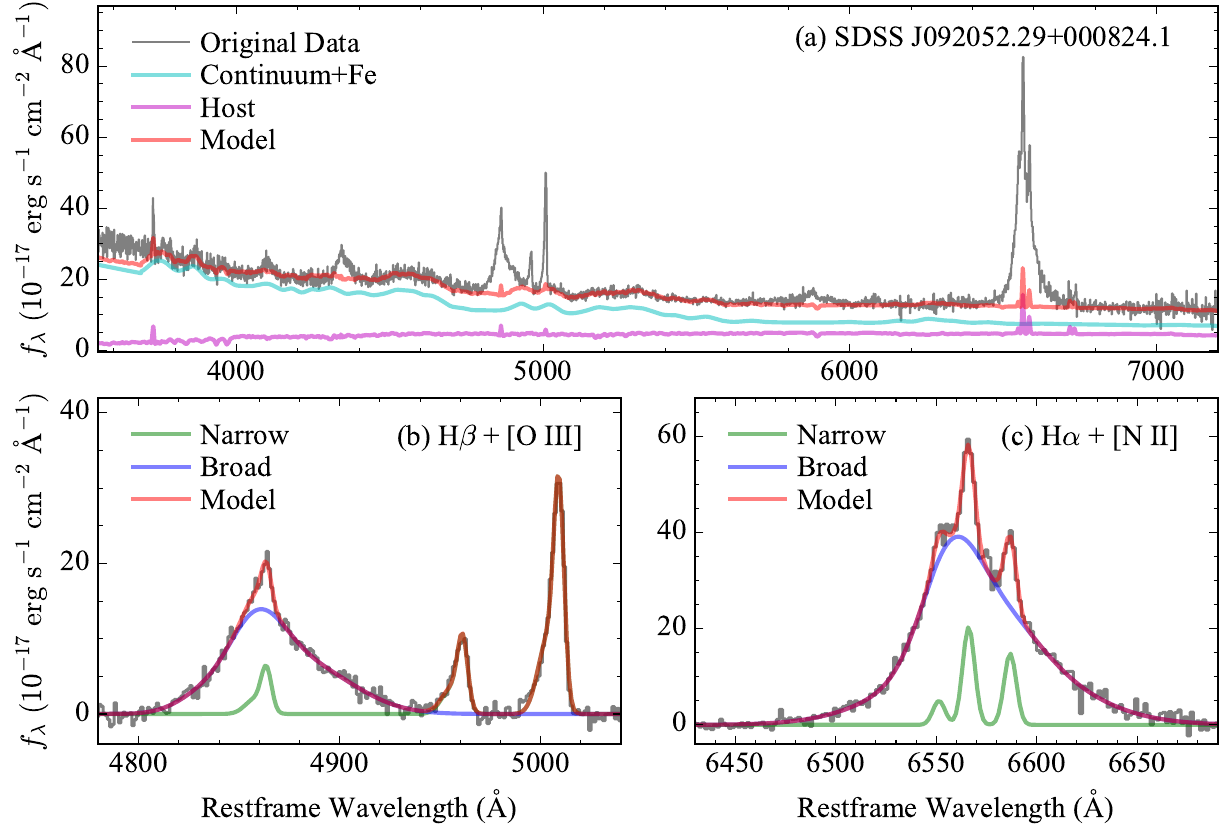}
\caption{Example of the spectral fitting. (a) The black solid line represents the original data in the rest-frame, corrected for the Milky Way reddening. The cyan line denotes the continuum from the accretion disk, modeled with a power law plus polynomial terms, along with \feii\ multiplets, while the magenta line shows the host galaxy component. The red line represents the best-fit model. Spectral decomposition of the (b) \hb\ and \oiii\ and (c) \hal\ and \nii\ region. The black histogram shows the residual spectrum after subtracting the continuum fit. The blue lines represent the broad components of \hb\ and \hal, while the green lines represent the narrow components of \hb\ and \hal, with the \oiii\ doublet and \nii, respectively. The red line indicates the total best-fit model. 
}
\end{figure*}

In particular, changing-look AGNs offer an opportunity to understand the physical origin of Balmer decrement variations in the BLR, as their values changed dramatically during a changing phase. The changing-look phenomenon is often thought to be linked to a change in the accretion state, rather than line-of-sight absorption or extinction \cite[e.g.,][]{macleod_2016,sheng_2017,yang_2018}, particularly in cases that exhibit repeated changes (\citealt{wang_2025} and references therein). This result suggests that the accretion state also plays a vital role in the variation of the Balmer decrement. On the other hand, \cite{panda_2024} found no clear systematic relationship between temporal variations of the Balmer decrement and AGN luminosity in a sample of 32 changing-look AGNs compiled from the literature. These conflicting results highlight the need for further systematic investigation.

To understand the physical origin of the Balmer decrement in the BLR, we investigate its temporal variation in response to changes in AGN properties in a large sample of type 1 AGNs. We analyze multi-epoch spectra from the SDSS, using the methods detailed in Section 2. We perform spectral fitting in Section 3, and in Section 4, we present the dependence of the Balmer decrement on AGN properties. We then discuss the physical implications of our findings in Section 5 and conclude in Section 6. This study adopts the cosmological parameters from the 2018 Planck results ($H_0=67.36 \pm 0.54$ km s$^{-1}$ Mpc$^{-1}$, $\Omega_\Lambda=0.6847 \pm 0.0073$, $\Omega_m = 0.3153 \pm 0.0073$; \citealt{planck_2020}).

\section{Data} \label{sec:style}
We begin by selecting type 1 AGNs from the SDSS Data Release (DR) 16 quasar catalog (\citealt{lyke_2020}). To investigate temporal changes in the Balmer decrement, we focus on targets with multi-epoch spectroscopic observations available within the SDSS dataset. Due to systematic discrepancies in wavelength coverage and sensitivity between the original SDSS spectrograph and the Baryon Oscillation Spectroscopic Survey (BOSS; \citealp{smee_2013}) spectrograph, we exclude data obtained with the original SDSS spectrograph. As the goal of this study is to examine the flux ratio between \hal\ and \hb, we apply a redshift criterion of $z \leq 0.45$ to ensure complete spectral coverage of the \hal\ line. We further refine our sample through visual inspection, eliminating targets with spectral artifacts near the \hb\ or \hal\ lines. This selection process results in a final sample of 1,403 type 1 AGNs, yielding 69,275 pairwise combinations formed from multiple observations per source. The majority of targets ($\sim 86 \%$ of the entire sample) exhibit only one pairwise combination composed of two spectra, while a minority of the targets have more than a thousand spectral pairs because they were repeatedly observed as part of the SDSS reverberation mapping project.

\begin{figure*}[tp!]
\centering
\includegraphics[width=0.49\textwidth]{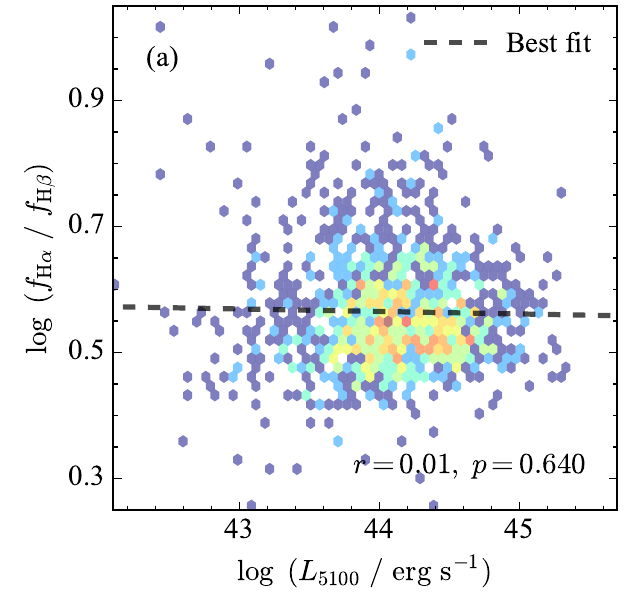}
\includegraphics[width=0.49\textwidth]{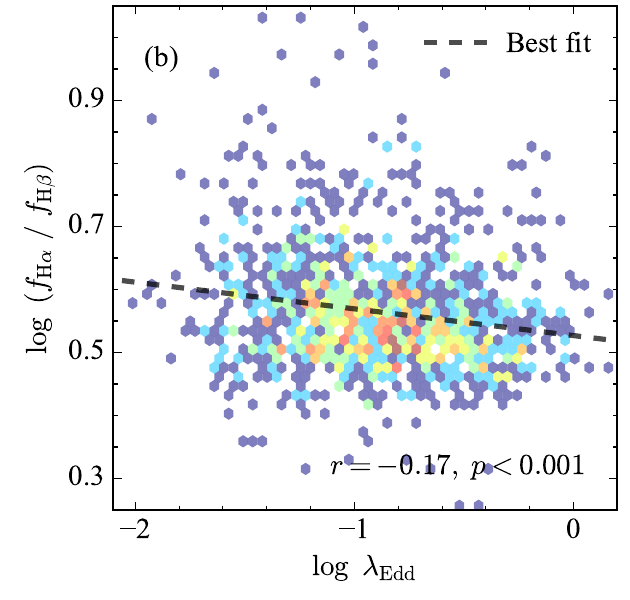}
\caption{Balmer decrement versus (a) continuum luminosity at 5100 \AA\ and (b) Eddington ratio. The dashed line shows the best-fit linear regression. The density contour is shown on a logarithmic scale.}
\end{figure*}

\hb\ and \hal\ emission lines are fitted from the continuum-subtracted spectra. For the \hb\ region, we adopt double Gaussian profiles for the narrow emission lines to account for the outflow component commonly seen in \oiii$\lambda\lambda 4959, 5007$ \cite[e.g.][]{greene_2005a,kim_2006}. 
\section{Spectral Fitting} 
In order to investigate the Balmer decrement in our sample, we perform spectral fitting on the BOSS spectra. We begin by correcting for Galactic extinction using the extinction law from \cite{fitzpatrick_1999} and $E(B-V)$ values adopted from \cite{schlafly_2011}. The extinction-corrected spectra are then shifted to the rest-frame using the redshift given in \citet{lyke_2020}, followed by continuum fitting to decompose the host galaxy component, \feii\ multiplets, and the power-law continuum to derive a pure emission-line spectrum. For this purpose, we use \texttt{PyQSOFit}, which was developed for spectral fitting of SDSS QSOs (\citealt{shen_2019}). We specifically adopt the method optimized for host decomposition (\citealt{ren_2024}), as this step is critical for robustly estimating the continuum emission from the accretion disk and the fluxes of the Balmer lines. The accretion disk continuum is modeled with a power-law and an additional polynomial function to account for occasional peculiarities in the continuum, possibly due to abnormal internal extinction \citep{shen_2019} or instrumental calibration \citep{greene_2006}. The \feii\ multiplets are fitted using the template from \cite{boroson_1992}, with the widths of the multiplets and their strengths treated as free parameters. The spectral area regions from strong emission lines are used for the fit.

The wavelength separation between \hb\ and \oiii\ is fixed to their known value. Additionally, the line ratio between two \oiii\ lines is fixed to the theoretical value of 2.98. This fitting process is performed in velocity space. It is worth noting that \sii$\lambda\lambda 6716, 6731$ is frequently adopted as a reference for narrow emission lines in the \hal\ + \nii$\lambda\lambda 6548, 6584$ region \citep{ho_1997b}. In our case, however, the signal-to-noise ratio (S/N) of \sii\ is generally too low for robust model fitting. For consistency, we therefore adopt the \oiii\ doublets as the reference for both the \hb\ and \hal\ regions. Visual inspection of the line-fitting results indicates that adopting two components to fit the narrow emissions in the spectral region of \hal\ and \nii\ occasionally leads to overestimation of the fluxes of narrow emissions, and hence underestimation of the flux of broad \hal\ emission, possibly due to the complexity of line profiles in the \hal$+$\nii\ region. Therefore, we used a single Gaussian component represented by the narrower Gaussian profile derived from the \oiii\ emission for the narrow \hal\ and \nii\ emissions. Since the flux contribution of the second Gaussian component relative to the broad emissions is typically $\sim4\%$, the impact of this choice is minimal.

For the broad components, we use three Gaussian components, as they are generally sufficient to yield an acceptable fit of the broad Balmer lines. When the peak of broad emission lines (i.e., \hb\ and \hal) is smaller than three times the uncertainty in the corresponding continuum flux density, we assume that the measurement of line fluxes is highly uncertain. Consequently, we exclude those objects to ensure the robustness of the flux estimation. Additionally, we visually inspect the fitting results and discard objects with poor fitting, primarily due to the residual from the imperfect sky subtraction or the lack of sufficient spectral elements in the \hb\ or \hal\ regions. The final sample used for further analysis contains 1,116 targets with 44,679 pairwise combinations from multiple spectra per source. While the majority of the targets ($\sim86\%$ of the entire sample) exhibit one pairwise combination composed of two spectra for each object, a minority of the targets have more than 2000 spectroscopic pairwise combinations, as those were repeatedly observed as a part of the RM SDSS project.

An example of spectral fitting is shown in Figure 1. Uncertainties in the spectral measurements, such as flux and full width at half maximum (FWHM), are computed using a bootstrap approach, in which the spectral fit is applied repeatedly to spectra with flux densities perturbed within their uncertainties.

\section{Results}
\subsection{Balmer Decrement vs. AGN Properties}

The mean of the distribution of the Balmer decrement ($f_{\rm H\alpha} / f_{\rm H\beta}$) is $3.66\pm0.95$, which is marginally larger than the value of 3.1 predicted from the Case B recombination under conditions appropriate for AGNs \citep{halpern_1982,halpern_1983,gaskell_1984}. The typical measurement error of the Balmer decrement is $\sim0.1$. We note that a small fraction ($\sim3\%$) of our sample has a Balmer decrement smaller than the prediction from the normal case B recombination ($2.7$). This is likely attributable to measurement errors or the systematic uncertainty arising from the imperfect decomposition between the broad and narrow emission lines. It is worthwhile to note that the average Balmer decrement of the narrow emission lines is $2.70$ with a standard deviation of $1.01$, which is closer to the prediction from the case B condition compared to the broad emissions. In order to examine whether the Balmer decrement correlates with AGN properties, we calculate the black hole (BH) mass ($M_{\rm BH}$), AGN bolometric luminosity ($L_{\rm bol}$), and Eddington ratio ($\lambda_{\rm Edd}$) directly from the spectral measurements. The $M_{\rm BH}$ is measured using the virial method applied to single-epoch spectra. Specifically, the estimated FWHM of the broad emission lines (\hal) and the monochromatic luminosity at $5100$ \AA\ ($L_{\rm 5100}$), together with the formalisms from \cite{greene_2005} and \cite{ho_2015}, are adopted. 
We derive $L_{\rm bol}$ from the $L_{\rm 5100}$, which is measured from the host galaxy-subtracted spectrum, adopting the bolometric correction ($L_{\rm bol} = 9.4 L_{\rm 5100}$) from \cite{richards_2006}. Finally, the $\lambda_{\rm Edd}$ is computed based on the bolometric luminosity and BH mass derived from the broad \hal\ emissions. The AGN properties of the final sample are $\log\ (M_{\mathrm{BH}}/M_{\odot}) = 7.85 \pm 0.54$, $\log\ (L_{\mathrm{bol}}/\mathrm{erg\,s^{-1}}) = 45.06 \pm 0.48$, and $\log \lambda_{\mathrm{Edd}} = -0.89 \pm 0.39$.

\begin{figure}[tp!]
\centering
\includegraphics[width=0.49\textwidth]{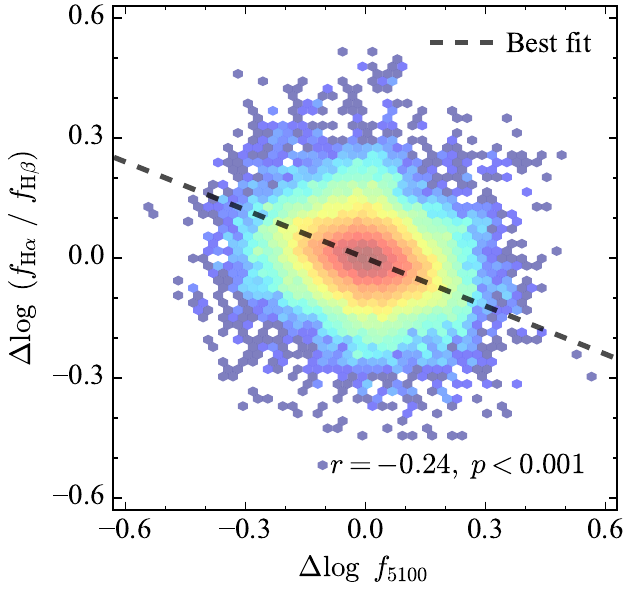}
\caption{Comparison of temporal variation of the Balmer decrement with AGN brightness indicated by the continuum flux at 5100 \AA. The density contour is shown on a logarithmic scale.}
\end{figure}


In this analysis, we use the mean values of the spectral measurements because multiple spectra were obtained from the same target. As the physical origin of the Balmer decrements remains uncertain, any correction for possible extinction due to the AGN is not applied in this study. We find that the Balmer decrement shows no correlation with $L_{5100}$ with a Spearman correlation coefficient ($r$) of 0.01, whereas it is anti-correlated with the Eddington ratio with $r=-0.17$ (Fig. 2). We note that all correlation coefficients derived in this study are statistically significant ($p$-value $\ll 0.001$), except for the relation between the Balmer decrement and $L_{5100}$, and therefore $p$-values are omitted hereafter. 

The lack of correlation between the Balmer decrement and AGN luminosity is consistent with previous findings \citep{dong_2008,lu_2019b}. The observed anti-correlation between the Balmer decrement and the Eddington ratio was also reported by \cite{wu_2023} ($r=-0.51$), although our measured correlation is much weaker ($r=-0.17$). The difference may reflect the fact that our sample is, on average, brighter than that of \cite{wu_2023}, which focused on low-luminosity AGNs. \cite{lu_2019b} also found a similar anti-correlation between Balmer decrement and dimensionless accretion rate ($\dot{\mathscr{M}}=\lambda_{\rm Edd}/\eta$) for SDSS DR7 AGNs, with $r=-0.18$, which is similar to our value.

\subsection{Temporal Variation of Balmer Decrement}
The primary aim of this study is to investigate the temporal variation of the Balmer decrement, $\Delta f_{\rm H\alpha} / f_{\rm H\beta}$, and its correlation with the temporal variation of AGN properties. For this purpose, we examine the temporal behavior of the Balmer decrement for each individual target. To reduce uncertainties from flux calibration, the spectra are additionally normalized by matching the fluxes of the \oiii\ $\lambda5007$ emission lines in each spectroscopic pair. 

Figure 3 compares the temporal change in the Balmer decrement with that of the AGN brightness, represented by the continuum flux at 5100 \AA. A negative correlation becomes evident ($r = -0.24$), in contrast with its absence in the mean spectra of the same sample (Section 4.1). While the physical origin of this discrepancy is unclear, it is worth noting that temporal variations in an individual object provide an excellent tool for studying the roles of AGN luminosity while the BH mass remains unchanged. An orthogonal distance regression fit yields $\Delta \log\ (f_{\rm H\alpha}/f_{  H\beta}) = -0.40 \Delta \log f_{\rm 5100}$. The zero-point of the correlation is close to zero, suggesting that the trend may be independent of the faint and bright phases of AGNs. We note that the temporal variation of AGN brightness corresponds to that of the Eddington ratio, since the BH mass is fixed for a given target. Therefore, the observed anti-correlation of the Balmer decrement variation with AGN brightness variation may reflect its anti-correlation with the Eddington ratio variation.

\section{Implications}
\subsection{Physical Origin of Balmer Decrement}
One of the striking findings of this study is that the temporal variation of the BLR Balmer decrement is primarily anti-correlated to that of AGN luminosity. A key question is whether this anti-correlation arises from dust extinction along the line-of-sight or from physical conditions within the BLR. To further examine the causal connection with extinction in AGNs, we estimate the variation of continuum color together with the Balmer decrement. Using the continuum solely from the accretion disk derived from the spectral fit, we measure the continuum fluxes at 3800 \AA\ ($f_{3800}$) and 5100 \AA\ ($f_{5100}$). 

\begin{figure}[tp!]
\centering
\includegraphics[width=0.49\textwidth]{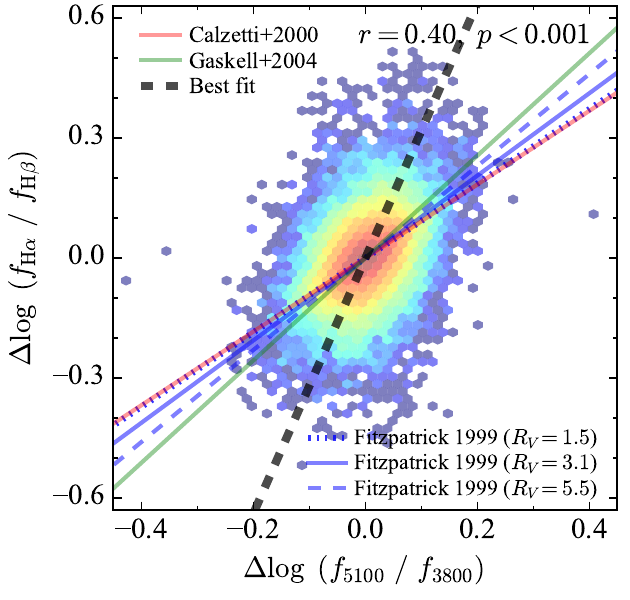}
\caption{Comparison of the temporal variation between the Balmer decrement and the flux ratio of the continua at 3800 and 5100 \AA. The various lines represent the temporal variation predicted solely by the intrinsic extinction of AGNs, under the assumption that the accretion disk and BLR share the same extinction. The blue dotted, solid, and dashed lines correspond to the predictions based on the extinction curve of \citet{fitzpatrick_1999} with $R_V = 1.5,3.1,$ and $5.5$, respectively. The red and green lines are derived from the extinction curves of star-forming galaxies (\citealt{calzetti_2000}) and AGNs (\citealt{gaskell_2004}), respectively. The black dashed line denotes the orthogonal distance regression of the observed data.}
\end{figure}

If the temporal variation of the BLR Balmer decrement is driven by extinction, and the BLR and accretion disk experience a similar amount of extinction along the line-of-sight, the continuum color and Balmer decrement should be strongly correlated, with a relation consistent with predictions from the extinction law. This test can only be carried out with the temporal changes in the same target, as different samples may have intrinsically different continuum slopes depending on the temperature of the accretion disk. Figure 4 shows that the temporal variation of the Balmer decrement is strongly correlated with that of the continuum color, suggesting that the extinction could be a physical origin of the Balmer decrement variation. Interestingly, however, the slope of this relation is significantly steeper than that predicted by dust extinction, as judged by the extinction curves of the Milky Way, star-forming galaxies, and AGNs, adopted from \cite{fitzpatrick_1999}, \cite{calzetti_2000}, and \cite{gaskell_2004}, respectively. 

In the case of the extinction curve of \cite{fitzpatrick_1999}, we attempt to reproduce the observed trend by varying the $R_V$ value ($R_V = 1.5,\,3.1,$ and $5.5$) to adjust the extinction curve slope, but cannot find parameters that match the observed relation between the Balmer decrement and continuum color variation. Notably, extinction curves derived from AGN composite spectra in previous studies differ significantly from those of ordinary galaxies \cite[e.g.,][]{czerny_2004, gaskell_2004}. \cite{gaskell_2004} argued that the AGN extinction curve is flat and wavelength-independent  (``gray''), whereas other studies reported that the reddening curve of individual AGNs is comparable to or steeper than that of the Milky Way and the Small Magellanic Cloud \cite[e.g.,][]{crenshaw_2001, crenshaw_2002}. However, even the AGN extinction curve of \cite{gaskell_2004} predicts a shallower relation between the temporal variation of the Balmer decrement and the continuum color than our observed relation. 

This suggests the extinction may not be the sole cause of Balmer decrement variability. However, caution is warranted, as continuum color also changes with AGN brightness (i.e., bluer-when-brighter; \citealt{giveon_1999, ruan_2014}). Given the negative correlation between the Balmer decrement and continuum flux, a decrease in AGN flux could produce both a steeper Balmer decrement and a redder continuum simultaneously, making it hard to disentangle the effect of reddening.

\begin{figure}[tp!]
\centering
\includegraphics[width=0.49\textwidth]{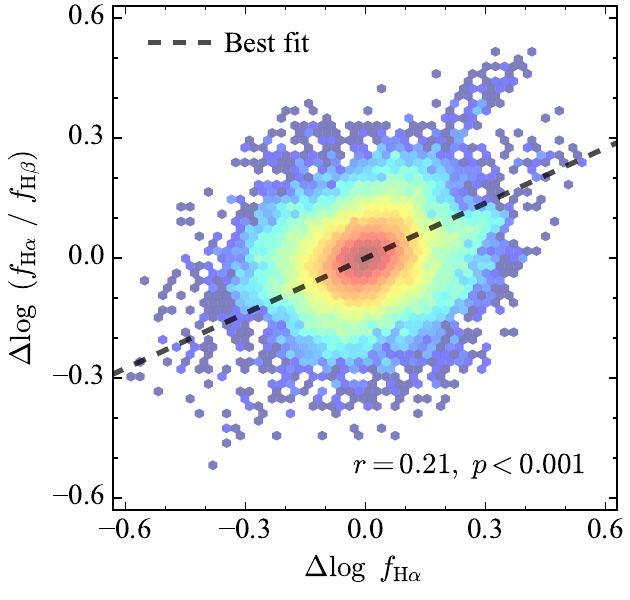}
\caption{Comparison of temporal variation between the Balmer decrement and AGN brightness indicated by the flux of broad \hal\ emission. The dashed line shows the best-fit linear regression. The density contour is shown on a logarithmic scale.}
\end{figure}

Alternatively, this discrepancy with the prediction from the extinction curve can be reconciled if the accretion disk and BLR do not share the same extinction. This scenario is plausible, as the dusty torus is clumpy and the BLR is significantly larger in extent than the accretion disk. Therefore, the amount of extinction can differ between the two components, depending on the geometry and covering factor of the clumpy torus. This effect would more likely introduce random scatter into the relation between the Balmer decrement and continuum color, rather than cause a systematic change (i.e., a significant shift in the slope). However, if the torus covering factor systematically increases in response to a decline in AGN luminosity or Eddington ratio, and if the extinction is more severe for the BLR than for the accretion disk, the observed result in Figure 4 can be produced. Interestingly, the former assumption is in broad agreement with the observed anti-correlation between torus covering factor and either luminosity \cite[e.g.,][]{maiolino_2007,lusso_2013,trefoloni_2025} or Eddington ratio \cite[for $0.01 \lesssim \log \lambda_{\rm Edd} \lesssim 0.3$; e.g.,][]{ricci_2017c,zhuang_2018,ricci_2023}. Although several studies have reported different results, showing that the covering factor is only weakly correlated or not correlated with AGN luminosity \cite[e.g.,][]{netzer_2016, stalevski_2016}, the scenario that the accretion disk and BLR do not share the same extinction cannot be completely ruled out.


\begin{figure*}[tp!]
\centering
\includegraphics[width=0.49\textwidth]{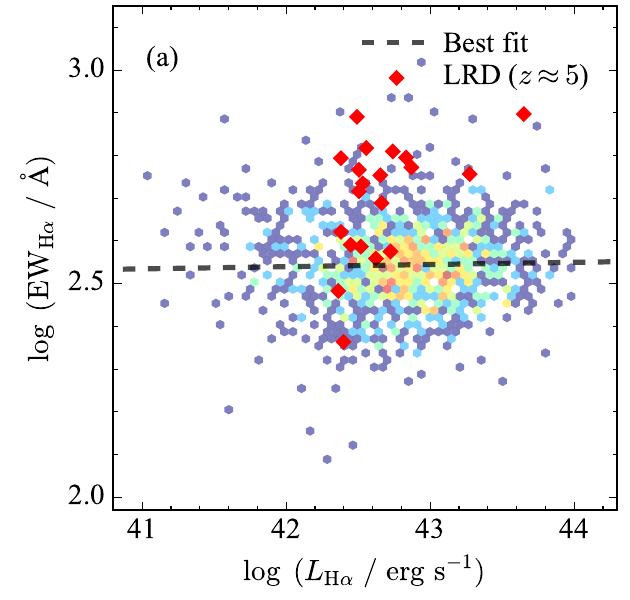}
\includegraphics[width=0.49\textwidth]{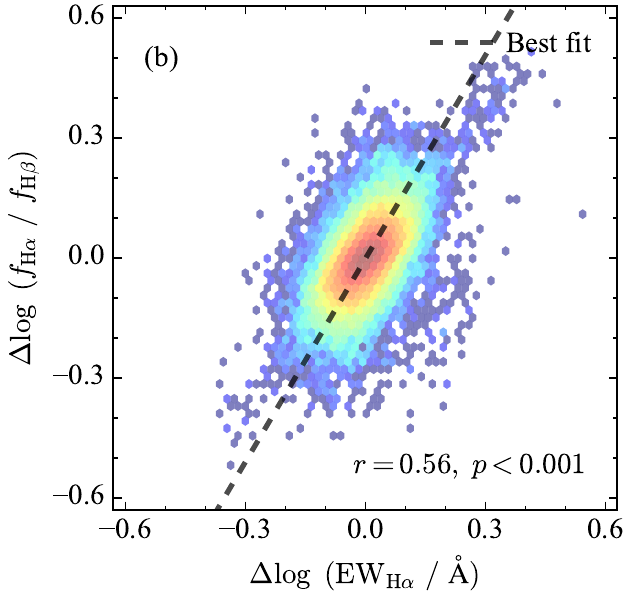}
\caption{(a) Comparison of the broad \hal\ luminosity with its equivalent width (EW). The red diamonds denote the little red dots at $z\approx5$ drawn from \citep{matthee_2024}. (b) Comparison of the temporal variation between the BLR Balmer decrement and the EW of the broad \hal. The dashed line shows the best-fit linear regression. The density contour is shown on a logarithmic scale.}
\end{figure*}

The more plausible explanation, rather than the reddening, is that the anti-correlation between the temporal variation of the Balmer decrement and that of the continuum luminosity is naturally driven by the difference in responsivities of \hal\ and \hb\ within the BLR, without invoking reddening. Responsivity, which describes how sensitively a line flux responds to continuum variation, is lower for \hal\ than for \hb\ due to its higher optical depth \cite[e.g.,][]{netzer_1975,rees_1989,korista_2004}. Under extreme BLR conditions of high density, the responsivities of \hal\ and \hb\ are strongly affected by optical depth, as Balmer transitions that are assumed to be optically thin in the Case B recombination become optically thick. In addition, collisional excitation, which enhances \hal\ more than \hb\ \cite[e.g.][]{gaskell_1984,osterbrock_2006}, can further contribute to lower responsivity of \hal\ relative to \hb. Therefore, when the AGN dims, the \hal\ flux decreases less than \hb, resulting in a steeper \hal/\hb\ ratio.

Similarly, previous spectral monitoring studies of individual type 1 \citep{popovic_2011} and changing-look AGNs \citep{shapovalova_2004,shapovalova_2010,kollatschny_2018} have reported that the temporal variation of the Balmer decrement anti-correlates with that of the continuum luminosity, which is consistent with findings from a multi-object spectral monitoring analysis \citep{ma_2023}. This anti-correlation also appears as the earlier disappearance of the broad \hb\ component than the broad \hal\ component during the transition to a type 2 state in changing-look AGNs \cite[e.g.,][]{kim_2018,li_2022}. In the changing-look AGN 1ES 1927+654, \citet{li_2022} found that the Balmer decrement dramatically increased by a factor of $\sim 4$–$5$ during the dimming phase over a timescale of $\sim 500$ days. They argued that this steepening was not due to changes in reddening but instead reflected decreases in both electron density and the ionizing photon flux. These arguments align well with our findings, supporting the conclusion that the observed anti-correlation between the Balmer decrement and continuum strength is not caused by a reddening along the line-of-sight, but is likely driven by BLR gas conditions. 

Remarkably, the temporal variation between the Balmer decrement and the \hal\ flux shows the opposite behavior, exhibiting a positive correlation (Fig. 5). It is worthwhile to note that the temporal variation of the \hb\ flux is anti-correlated with the Balmer decrement. This trend is similar to the observed relation for the continuum luminosity (Fig. 3), and has also been reported for individual AGNs \cite[e.g.,][]{kollatschny_2000,kollatschny_2022}. The \hal\ luminosity is often used as a tracer for the continuum luminosity, such as $L_{5100}$, based on the observed linear relation between the two parameters \cite[e.g.,][]{greene_2005}. Under this assumption, the BH mass has been estimated using the \hal\ luminosity in combination with the FWHM of \hal. Therefore, the observed positive correlation between the temporal Balmer decrement and the \hal\ flux conflicts with our finding that the temporal Balmer decrement is anti-correlated with the continuum flux. This discrepancy may arise from the complicated interplay of radiative transfer effects and excitation mechanisms in the BLR, and suggests that the \hal\ luminosity is correlated with both $L_{5100}$ {\it and} the Balmer decrement, according to the relation 
\begin{equation}
\log L_{5100}  = 0.99 \log L_{{\rm H}\alpha} -0.48 \log\ (f_{\rm H \alpha}/f_{\rm H \beta}) + 1.98, 
\end{equation}
where luminosities are expressed in units of erg s$^{-1}$. The scatter in the relation is $\sim0.11$ dex. This relation can be useful for inferring $L_{5100}$ from luminosity measurements of the broad emission lines. Interestingly, this effect results in the strong correlation of the equivalent width (EW) of \hal\ with the Balmer decrement (Fig. 6). 

In addition to the temporal behavior, we also found that the Balmer decrement is inversely correlated with the Eddington ratio in the mean spectra. \citet{wu_2023} reported that, at low accretion rates, the Balmer decrement increases sharply, which they attributed to a dramatic reduction in ultraviolet photons as the accretion state transitions to a truncated disk model \citealt{ho_2008}). In our analysis based on the mean spectra, a dependence on the Eddington ratio is very weak, possibly because \citet{wu_2023} concentrated on AGNs with lower Eddington ratios reaching as low as $\log \lambda_{\rm Edd} \sim -3$.
Nevertheless, their conclusion that the Balmer decrement is primarily regulated by the strength of ultraviolet photons remains consistent with our finding that the temporal variation of the Balmer decrement is anti-correlated with that of AGN luminosity. 

\subsection{Application to high-z AGNs}
The James Webb Space Telescope (JWST) has opened a new window on the demographics of AGNs in the early Universe through the discovery of a newly identified AGN population that appears red and compact, the so-called little red dots (LRDs; e.g., \citealt{kocevski_2023, maiolino_2024}). These objects exhibit a V-shaped spectral energy distribution and are exceptionally common at high redshift \cite[e.g.,][]{harikane_2023, greene_2024, kokorev_2024}, although their physical origin remains under debate \cite[e.g.,][]{inayoshi_2025, jiang_2025}. One of the remarkable features of LRDs is the relatively high EW of the broad \hal\ emission, associated with a steep Balmer decrement \cite[e.g.,][]{matthee_2024, brooks_2025, Maiolino_2025}. This finding may suggest that the two parameters are physically related. To investigate whether this relation is also present in our sample, we estimate the EW of the broad \hal\ using the host-subtracted continuum. Figure 6a shows that high-$z$ LRDs from \citet{matthee_2024} tend to have higher EWs than those in our sample. It should be noted that the EW of LRDs is estimated without subtracting the host contribution. Therefore, their measurements should be regarded as lower limits.

To assess the causal connection between the steep Balmer decrement and the EW of \hal, which is clearly observed in high-$z$ AGNs, we examined the temporal variation of these two parameters in our sample. Remarkably, their variations are strongly correlated, and the trend is in excellent agreement with that seen in high-$z$ LRDs (i.e., higher \hal\ EW with a steeper Balmer decrement; Fig. 6b). This finding suggests that the two features are closely linked to the physical conditions in LRDs, and that distinctive radiative transfer effects and excitation mechanisms within the BLR, rather than reddening, may be responsible for the steep Balmer decrement in LRDs.

\section{Summary}
To unveil the physical origin of the Balmer decrement in the broad-line region (BLR) of type 1 AGNs,  we utilize the multi-epoch spectroscopic dataset of more than 1,000 low-redshift AGNs from the SDSS. Performing detailed spectra fitting to derive the Balmer decrement of the broad lines, we arrive at the following conclusions:  

\begin{itemize}
\item{The analysis based on the mean spectra reveals that the Balmer decrement is not correlated with the AGN luminosity, but is marginally inversely correlated with the Eddington ratio. These results are broadly consistent with the previous studies based on the mean spectra.}

\item{Unlike the result from the mean spectra, the analysis of the spectra of individual objects reveals that the temporal variation of the Balmer decrement is inversely correlated with the AGN luminosity, steepening as the AGN brightens. The slope and correlation coefficient of the relation are larger than those from the mean spectra, indicating that the change of the brightness plays a critical role in determining the Balmer decrement.}

\item{A comparison of the variations in continuum color and the Balmer decrement reveals that the nominal dust extinction curves, even in extreme cases, fail to reproduce the observed trend, suggesting that intrinsic reddening in AGNs is unlikely to be a dominant factor in determining the variation of the Balmer decrement.}

\item{The Balmer decrement is likely governed by the optical depth effect and collisional excitation arising from the high density of the BLR. In particular, differences in the responsivities of \hal\ and \hb\ to the ionizing continuum may play an important role in the variation of the Balmer decrement.}

\end{itemize}
Our study reveals that the variation of the Balmer decrement is predominantly determined by the physical conditions within the BLR, rather than by reddening along the line-of-sight. This interpretation provides useful insight into the underlying mechanisms not only of changing-look AGNs but also of the steep Balmer decrements commonly seen in high-redshift AGNs \cite[e.g.,][]{brooks_2025}.

\begin{acknowledgments}
We are grateful to the anonymous referee for the constructive comments, which greatly improved the manuscript. LCH was supported by the National Science Foundation of China (12233001) and the China Manned Space Program (CMS-CSST-2025-A09). This work was supported by the National Research Foundation of Korea (NRF) grant funded by the Korean government (MSIT) (Nos. RS-2024-00347548 and RS-2025-16066624). 
\end{acknowledgments}

\facilities{Sloan}

\software{astropy \citep{2013A&A...558A..33A,2018AJ....156..123A,2022ApJ...935..167A},  
          }

\bibliography{torus}{}

@ARTICLE{ho_2015,
   author = {{Ho}, L.~C. and {Kim}, M.},
    title = "{A Revised Calibration of the Virial Mass Estimator for Black Holes in Active Galaxies Based on Single-epoch H{$\beta$} Spectra}",
  journal = {\apj},
archivePrefix = "arXiv",
   eprint = {1507.04821},
     year = 2015,
    month = aug,
   volume = 809,
      eid = {123},
    pages = {123},
      doi = {10.1088/0004-637X/809/2/123},
   adsurl = {http://adsabs.harvard.edu/abs/2015ApJ...809..123H}
}

@ARTICLE{kim_2006,
   author = {{Kim}, M. and {Ho}, L.~C. and {Im}, M.},
    title = "{Constraints on the Star Formation Rate in Active Galaxies}",
  journal = {\apj},
   eprint = {astro-ph/0601316},
     year = 2006,
    month = may,
   volume = 642,
    pages = {702-710},
      doi = {10.1086/501422},
   adsurl = {http://adsabs.harvard.edu/abs/2006ApJ...642..702K}
}

@ARTICLE{schlafly_2011,
       author = {{Schlafly}, Edward F. and {Finkbeiner}, Douglas P.},
        title = "{Measuring Reddening with Sloan Digital Sky Survey Stellar Spectra and Recalibrating SFD}",
      journal = {\apj},
         year = 2011,
        month = aug,
       volume = {737},
       number = {2},
          eid = {103},
        pages = {103},
          doi = {10.1088/0004-637X/737/2/103},
archivePrefix = {arXiv},
       eprint = {1012.4804},
 primaryClass = {astro-ph.GA},
       adsurl = {https://ui.adsabs.harvard.edu/abs/2011ApJ...737..103S}
}

@ARTICLE{greene_2005,
       author = {{Greene}, Jenny E. and {Ho}, Luis C.},
        title = "{Estimating Black Hole Masses in Active Galaxies Using the H{\ensuremath{\alpha}} Emission Line}",
      journal = {\apj},
         year = 2005,
        month = sep,
       volume = {630},
       number = {1},
        pages = {122-129},
          doi = {10.1086/431897},
archivePrefix = {arXiv},
       eprint = {astro-ph/0508335},
 primaryClass = {astro-ph},
       adsurl = {https://ui.adsabs.harvard.edu/abs/2005ApJ...630..122G}
}

@ARTICLE{greene_2005a,
       author = {{Greene}, Jenny E. and {Ho}, Luis C.},
        title = "{A Comparison of Stellar and Gaseous Kinematics in the Nuclei of Active Galaxies}",
      journal = {\apj},
         year = 2005,
        month = jul,
       volume = {627},
       number = {2},
        pages = {721-732},
          doi = {10.1086/430590},
archivePrefix = {arXiv},
       eprint = {astro-ph/0503675},
 primaryClass = {astro-ph},
       adsurl = {https://ui.adsabs.harvard.edu/abs/2005ApJ...627..721G}
}

@ARTICLE{zhuang_2018,
       author = {{Zhuang}, Ming-Yang and {Ho}, Luis C. and {Shangguan}, Jinyi},
        title = "{The Infrared Emission and Opening Angle of the Torus in Quasars}",
      journal = {\apj},
         year = 2018,
        month = aug,
       volume = {862},
       number = {2},
          eid = {118},
        pages = {118},
          doi = {10.3847/1538-4357/aacc2d},
archivePrefix = {arXiv},
       eprint = {1806.03783},
 primaryClass = {astro-ph.GA},
       adsurl = {https://ui.adsabs.harvard.edu/abs/2018ApJ...862..118Z}
}

@ARTICLE{ricci_2017c,
       author = {{Ricci}, Claudio and {Trakhtenbrot}, Benny and {Koss}, Michael J. and {Ueda}, Yoshihiro and {Schawinski}, Kevin and {Oh}, Kyuseok and {Lamperti}, Isabella and {Mushotzky}, Richard and {Treister}, Ezequiel and {Ho}, Luis C. and {Weigel}, Anna and {Bauer}, Franz E. and {Paltani}, Stephane and {Fabian}, Andrew C. and {Xie}, Yanxia and {Gehrels}, Neil},
        title = "{The close environments of accreting massive black holes are shaped by radiative feedback}",
      journal = {\nat},
         year = 2017,
        month = sep,
       volume = {549},
       number = {7673},
        pages = {488-491},
          doi = {10.1038/nature23906},
archivePrefix = {arXiv},
       eprint = {1709.09651},
 primaryClass = {astro-ph.HE},
       adsurl = {https://ui.adsabs.harvard.edu/abs/2017Natur.549..488R}
}

@ARTICLE{netzer_2016,
       author = {{Netzer}, Hagai and {Lani}, Caterina and {Nordon}, Raanan and {Trakhtenbrot}, Benny and {Lira}, Paulina and {Shemmer}, Ohad},
        title = "{Star Formation Black Hole Growth and Dusty Tori in the Most Luminous AGNs at z=2-3.5}",
      journal = {\apj},
         year = 2016,
        month = mar,
       volume = {819},
       number = {2},
          eid = {123},
        pages = {123},
          doi = {10.3847/0004-637X/819/2/123},
archivePrefix = {arXiv},
       eprint = {1511.07876},
 primaryClass = {astro-ph.GA},
       adsurl = {https://ui.adsabs.harvard.edu/abs/2016ApJ...819..123N}
}

@ARTICLE{maiolino_2007,
       author = {{Maiolino}, R. and {Shemmer}, O. and {Imanishi}, M. and {Netzer}, H. and {Oliva}, E. and {Lutz}, D. and {Sturm}, E.},
        title = "{Dust covering factor, silicate emission, and star formation in luminous QSOs}",
      journal = {\aap},
         year = 2007,
        month = jun,
       volume = {468},
       number = {3},
        pages = {979-992},
          doi = {10.1051/0004-6361:20077252},
archivePrefix = {arXiv},
       eprint = {0704.1559},
 primaryClass = {astro-ph},
       adsurl = {https://ui.adsabs.harvard.edu/abs/2007A&A...468..979M}
}

@ARTICLE{mor_2011,
       author = {{Mor}, Rivay and {Trakhtenbrot}, Benny},
        title = "{Hot-dust Clouds with Pure-graphite Composition around type-I Active Galactic Nuclei}",
      journal = {\apjl},
         year = 2011,
        month = aug,
       volume = {737},
       number = {2},
        pages = {L36},
          doi = {10.1088/2041-8205/737/2/L36},
archivePrefix = {arXiv},
       eprint = {1105.3198},
 primaryClass = {astro-ph.CO},
       adsurl = {https://ui.adsabs.harvard.edu/abs/2011ApJ...737L..36M}
}

@ARTICLE{lusso_2013,
       author = {{Lusso}, E. and {Hennawi}, J.~F. and {Comastri}, A. and {Zamorani}, G. and {Richards}, G.~T. and {Vignali}, C. and {Treister}, E. and {Schawinski}, K. and {Salvato}, M. and {Gilli}, R.},
        title = "{The Obscured Fraction of Active Galactic Nuclei in the XMM-COSMOS Survey: A Spectral Energy Distribution Perspective}",
      journal = {\apj},
         year = 2013,
        month = nov,
       volume = {777},
       number = {2},
          eid = {86},
        pages = {86},
          doi = {10.1088/0004-637X/777/2/86},
archivePrefix = {arXiv},
       eprint = {1309.0814},
 primaryClass = {astro-ph.CO},
       adsurl = {https://ui.adsabs.harvard.edu/abs/2013ApJ...777...86L}
}

@ARTICLE{richards_2006,
       author = {{Richards}, Gordon T. and {Lacy}, Mark and {Storrie-Lombardi}, Lisa J. and {Hall}, Patrick B. and {Gallagher}, S.~C. and {Hines}, Dean C. and {Fan}, Xiaohui and {Papovich}, Casey and {Vanden Berk}, Daniel E. and {Trammell}, George B. and {Schneider}, Donald P. and {Vestergaard}, Marianne and {York}, Donald G. and {Jester}, Sebastian and {Anderson}, Scott F. and {Budav{\'a}ri}, Tam{\'a}s and {Szalay}, Alexander S.},
        title = "{Spectral Energy Distributions and Multiwavelength Selection of Type 1 Quasars}",
      journal = {\apjs},
         year = 2006,
        month = oct,
       volume = {166},
       number = {2},
        pages = {470-497},
          doi = {10.1086/506525},
archivePrefix = {arXiv},
       eprint = {astro-ph/0601558},
 primaryClass = {astro-ph},
       adsurl = {https://ui.adsabs.harvard.edu/abs/2006ApJS..166..470R}
}

@ARTICLE{yang_2018,
       author = {{Yang}, Qian and {Wu}, Xue-Bing and {Fan}, Xiaohui and {Jiang}, Linhua and {McGreer}, Ian and {Shangguan}, Jinyi and {Yao}, Su and {Wang}, Bingquan and {Joshi}, Ravi and {Green}, Richard and {Wang}, Feige and {Feng}, Xiaotong and {Fu}, Yuming and {Yang}, Jinyi and {Liu}, Yuanqi},
        title = "{Discovery of 21 New Changing-look AGNs in the Northern Sky}",
      journal = {\apj},
         year = 2018,
        month = aug,
       volume = {862},
       number = {2},
          eid = {109},
        pages = {109},
          doi = {10.3847/1538-4357/aaca3a},
archivePrefix = {arXiv},
       eprint = {1711.08122},
 primaryClass = {astro-ph.GA},
       adsurl = {https://ui.adsabs.harvard.edu/abs/2018ApJ...862..109Y}
}

@ARTICLE{fitzpatrick_1999,
       author = {{Fitzpatrick}, Edward L.},
        title = "{Correcting for the Effects of Interstellar Extinction}",
      journal = {\pasp},
         year = 1999,
        month = jan,
       volume = {111},
       number = {755},
        pages = {63-75},
          doi = {10.1086/316293},
archivePrefix = {arXiv},
       eprint = {astro-ph/9809387},
 primaryClass = {astro-ph},
       adsurl = {https://ui.adsabs.harvard.edu/abs/1999PASP..111...63F}
}

@ARTICLE{boroson_1992,
       author = {{Boroson}, Todd A. and {Green}, Richard F.},
        title = "{The Emission-Line Properties of Low-Redshift Quasi-stellar Objects}",
      journal = {\apjs},
         year = 1992,
        month = may,
       volume = {80},
        pages = {109},
          doi = {10.1086/191661},
       adsurl = {https://ui.adsabs.harvard.edu/abs/1992ApJS...80..109B}
}

@ARTICLE{planck_2020,
       author = {{Planck Collaboration} and {Aghanim}, N. and {Akrami}, Y. and {Ashdown}, M. and {Aumont}, J. and {Baccigalupi}, C. and {Ballardini}, M. and {Banday}, A.~J. and {Barreiro}, R.~B. and {Bartolo}, N. and {Basak}, S. and {Battye}, R. and {Benabed}, K. and {Bernard}, J. -P. and {Bersanelli}, M. and {Bielewicz}, P. and {Bock}, J.~J. and {Bond}, J.~R. and {Borrill}, J. and {Bouchet}, F.~R. and {Boulanger}, F. and {Bucher}, M. and {Burigana}, C. and {Butler}, R.~C. and {Calabrese}, E. and {Cardoso}, J. -F. and {Carron}, J. and {Challinor}, A. and {Chiang}, H.~C. and {Chluba}, J. and {Colombo}, L.~P.~L. and {Combet}, C. and {Contreras}, D. and {Crill}, B.~P. and {Cuttaia}, F. and {de Bernardis}, P. and {de Zotti}, G. and {Delabrouille}, J. and {Delouis}, J. -M. and {Di Valentino}, E. and {Diego}, J.~M. and {Dor{\'e}}, O. and {Douspis}, M. and {Ducout}, A. and {Dupac}, X. and {Dusini}, S. and {Efstathiou}, G. and {Elsner}, F. and {En{\ss}lin}, T.~A. and {Eriksen}, H.~K. and {Fantaye}, Y. and {Farhang}, M. and {Fergusson}, J. and {Fernandez-Cobos}, R. and {Finelli}, F. and {Forastieri}, F. and {Frailis}, M. and {Fraisse}, A.~A. and {Franceschi}, E. and {Frolov}, A. and {Galeotta}, S. and {Galli}, S. and {Ganga}, K. and {G{\'e}nova-Santos}, R.~T. and {Gerbino}, M. and {Ghosh}, T. and {Gonz{\'a}lez-Nuevo}, J. and {G{\'o}rski}, K.~M. and {Gratton}, S. and {Gruppuso}, A. and {Gudmundsson}, J.~E. and {Hamann}, J. and {Handley}, W. and {Hansen}, F.~K. and {Herranz}, D. and {Hildebrandt}, S.~R. and {Hivon}, E. and {Huang}, Z. and {Jaffe}, A.~H. and {Jones}, W.~C. and {Karakci}, A. and {Keih{\"a}nen}, E. and {Keskitalo}, R. and {Kiiveri}, K. and {Kim}, J. and {Kisner}, T.~S. and {Knox}, L. and {Krachmalnicoff}, N. and {Kunz}, M. and {Kurki-Suonio}, H. and {Lagache}, G. and {Lamarre}, J. -M. and {Lasenby}, A. and {Lattanzi}, M. and {Lawrence}, C.~R. and {Le Jeune}, M. and {Lemos}, P. and {Lesgourgues}, J. and {Levrier}, F. and {Lewis}, A. and {Liguori}, M. and {Lilje}, P.~B. and {Lilley}, M. and {Lindholm}, V. and {L{\'o}pez-Caniego}, M. and {Lubin}, P.~M. and {Ma}, Y. -Z. and {Mac{\'\i}as-P{\'e}rez}, J.~F. and {Maggio}, G. and {Maino}, D. and {Mandolesi}, N. and {Mangilli}, A. and {Marcos-Caballero}, A. and {Maris}, M. and {Martin}, P.~G. and {Martinelli}, M. and {Mart{\'\i}nez-Gonz{\'a}lez}, E. and {Matarrese}, S. and {Mauri}, N. and {McEwen}, J.~D. and {Meinhold}, P.~R. and {Melchiorri}, A. and {Mennella}, A. and {Migliaccio}, M. and {Millea}, M. and {Mitra}, S. and {Miville-Desch{\^e}nes}, M. -A. and {Molinari}, D. and {Montier}, L. and {Morgante}, G. and {Moss}, A. and {Natoli}, P. and {N{\o}rgaard-Nielsen}, H.~U. and {Pagano}, L. and {Paoletti}, D. and {Partridge}, B. and {Patanchon}, G. and {Peiris}, H.~V. and {Perrotta}, F. and {Pettorino}, V. and {Piacentini}, F. and {Polastri}, L. and {Polenta}, G. and {Puget}, J. -L. and {Rachen}, J.~P. and {Reinecke}, M. and {Remazeilles}, M. and {Renzi}, A. and {Rocha}, G. and {Rosset}, C. and {Roudier}, G. and {Rubi{\~n}o-Mart{\'\i}n}, J.~A. and {Ruiz-Granados}, B. and {Salvati}, L. and {Sandri}, M. and {Savelainen}, M. and {Scott}, D. and {Shellard}, E.~P.~S. and {Sirignano}, C. and {Sirri}, G. and {Spencer}, L.~D. and {Sunyaev}, R. and {Suur-Uski}, A. -S. and {Tauber}, J.~A. and {Tavagnacco}, D. and {Tenti}, M. and {Toffolatti}, L. and {Tomasi}, M. and {Trombetti}, T. and {Valenziano}, L. and {Valiviita}, J. and {Van Tent}, B. and {Vibert}, L. and {Vielva}, P. and {Villa}, F. and {Vittorio}, N. and {Wandelt}, B.~D. and {Wehus}, I.~K. and {White}, M. and {White}, S.~D.~M. and {Zacchei}, A. and {Zonca}, A.},
        title = "{Planck 2018 results. VI. Cosmological parameters}",
      journal = {\aap},
         year = 2020,
        month = sep,
       volume = {641},
          eid = {A6},
        pages = {A6},
          doi = {10.1051/0004-6361/201833910},
archivePrefix = {arXiv},
       eprint = {1807.06209},
 primaryClass = {astro-ph.CO},
       adsurl = {https://ui.adsabs.harvard.edu/abs/2020A&A...641A...6P}
}

@ARTICLE{ho_2008,
       author = {{Ho}, L.~C.},
        title = "{Nuclear activity in nearby galaxies.}",
      journal = {\araa},
         year = 2008,
        month = sep,
       volume = {46},
        pages = {475-539},
          doi = {10.1146/annurev.astro.45.051806.110546},
archivePrefix = {arXiv},
       eprint = {0803.2268},
 primaryClass = {astro-ph},
       adsurl = {https://ui.adsabs.harvard.edu/abs/2008ARA&A..46..475H}
}

@ARTICLE{ma_2023,
       author = {{Ma}, Yan-Song and {Li}, Shao-Jun and {Gu}, Chen-Sheng and {Jiang}, Jian-Xia and {Hou}, Kai-Li and {Qin}, Shu-Hao and {Bian}, Wei-Hao},
        title = "{The variability of the broad-line Balmer decrement for quasars from the Sloan Digital Sky Survey reverberation mapping}",
      journal = {\mnras},
         year = 2023,
        month = jul,
       volume = {522},
       number = {4},
        pages = {5680-5689},
          doi = {10.1093/mnras/stad1377},
archivePrefix = {arXiv},
       eprint = {2305.04637},
 primaryClass = {astro-ph.GA},
       adsurl = {https://ui.adsabs.harvard.edu/abs/2023MNRAS.522.5680M}
}

@ARTICLE{ruan_2014,
       author = {{Ruan}, John J. and {Anderson}, Scott F. and {Dexter}, Jason and {Agol}, Eric},
        title = "{Evidence for Large Temperature Fluctuations in Quasar Accretion Disks from Spectral Variability}",
      journal = {\apj},
         year = 2014,
        month = mar,
       volume = {783},
       number = {2},
          eid = {105},
        pages = {105},
          doi = {10.1088/0004-637X/783/2/105},
archivePrefix = {arXiv},
       eprint = {1401.1211},
 primaryClass = {astro-ph.HE},
       adsurl = {https://ui.adsabs.harvard.edu/abs/2014ApJ...783..105R}
}

@ARTICLE{wu_2023,
       author = {{Wu}, Jiancheng and {Wu}, Qingwen and {Xue}, Hanrui and {Lei}, Weihua and {Lyu}, Bing},
        title = "{Steep Balmer Decrement in Weak AGNs May Not Be Caused by Dust Extinction: Clues from Low-luminosity AGNs and Changing-look AGNs}",
      journal = {\apj},
         year = 2023,
        month = jun,
       volume = {950},
       number = {2},
          eid = {106},
        pages = {106},
          doi = {10.3847/1538-4357/acce9e},
archivePrefix = {arXiv},
       eprint = {2304.09435},
 primaryClass = {astro-ph.GA},
       adsurl = {https://ui.adsabs.harvard.edu/abs/2023ApJ...950..106W}
}

@ARTICLE{lyke_2020,
       author = {{Lyke}, Brad W. and {Higley}, Alexandra N. and {McLane}, J.~N. and {Schurhammer}, Danielle P. and {Myers}, Adam D. and {Ross}, Ashley J. and {Dawson}, Kyle and {Chabanier}, Sol{\`e}ne and {Martini}, Paul and {Busca}, Nicol{\'a}s G. and {Mas des Bourboux}, H{\'e}lion du and {Salvato}, Mara and {Streblyanska}, Alina and {Zarrouk}, Pauline and {Burtin}, Etienne and {Anderson}, Scott F. and {Bautista}, Julian and {Bizyaev}, Dmitry and {Brandt}, W.~N. and {Brinkmann}, Jonathan and {Brownstein}, Joel R. and {Comparat}, Johan and {Green}, Paul and {de la Macorra}, Axel and {Mu{\~n}oz Guti{\'e}rrez}, Andrea and {Hou}, Jiamin and {Newman}, Jeffrey A. and {Palanque-Delabrouille}, Nathalie and {P{\^a}ris}, Isabelle and {Percival}, Will J. and {Petitjean}, Patrick and {Rich}, James and {Rossi}, Graziano and {Schneider}, Donald P. and {Smith}, Alexander and {Vivek}, M. and {Weaver}, Benjamin Alan},
        title = "{The Sloan Digital Sky Survey Quasar Catalog: Sixteenth Data Release}",
      journal = {\apjs},
         year = 2020,
        month = sep,
       volume = {250},
       number = {1},
          eid = {8},
        pages = {8},
          doi = {10.3847/1538-4365/aba623},
archivePrefix = {arXiv},
       eprint = {2007.09001},
 primaryClass = {astro-ph.GA},
       adsurl = {https://ui.adsabs.harvard.edu/abs/2020ApJS..250....8L}
}

@ARTICLE{ren_2024,
       author = {{Ren}, Wenke and {Guo}, Hengxiao and {Shen}, Yue and {Silverman}, John D. and {Burke}, Colin J. and {Wang}, Shu and {Wang}, Junxian},
        title = "{Prior-informed Active Galactic Nucleus Host Spectral Decomposition Using PyQSOFit}",
      journal = {\apj},
         year = 2024,
        month = oct,
       volume = {974},
       number = {2},
          eid = {153},
        pages = {153},
          doi = {10.3847/1538-4357/ad6e76},
archivePrefix = {arXiv},
       eprint = {2406.17598},
 primaryClass = {astro-ph.GA},
       adsurl = {https://ui.adsabs.harvard.edu/abs/2024ApJ...974..153R}
}

@ARTICLE{shen_2019,
       author = {{Shen}, Yue and {Hall}, Patrick B. and {Horne}, Keith and {Zhu}, Guangtun and {McGreer}, Ian and {Simm}, Torben and {Trump}, Jonathan R. and {Kinemuchi}, Karen and {Brandt}, W.~N. and {Green}, Paul J. and {Grier}, C.~J. and {Guo}, Hengxiao and {Ho}, Luis C. and {Homayouni}, Yasaman and {Jiang}, Linhua and {I-Hsiu Li}, Jennifer and {Morganson}, Eric and {Petitjean}, Patrick and {Richards}, Gordon T. and {Schneider}, Donald P. and {Starkey}, D.~A. and {Wang}, Shu and {Chambers}, Ken and {Kaiser}, Nick and {Kudritzki}, Rolf-Peter and {Magnier}, Eugene and {Waters}, Christopher},
        title = "{The Sloan Digital Sky Survey Reverberation Mapping Project: Sample Characterization}",
      journal = {\apjs},
         year = 2019,
        month = apr,
       volume = {241},
       number = {2},
          eid = {34},
        pages = {34},
          doi = {10.3847/1538-4365/ab074f},
archivePrefix = {arXiv},
       eprint = {1810.01447},
 primaryClass = {astro-ph.GA},
       adsurl = {https://ui.adsabs.harvard.edu/abs/2019ApJS..241...34S}
}

@ARTICLE{czerny_2004,
       author = {{Czerny}, B. and {Li}, J. and {Loska}, Z. and {Szczerba}, R.},
        title = "{Extinction due to amorphous carbon grains in red quasars from the Sloan Digital Sky Survey}",
      journal = {\mnras},
         year = 2004,
        month = mar,
       volume = {348},
       number = {3},
        pages = {L54-L57},
          doi = {10.1111/j.1365-2966.2004.07590.x},
archivePrefix = {arXiv},
       eprint = {astro-ph/0401158},
 primaryClass = {astro-ph},
       adsurl = {https://ui.adsabs.harvard.edu/abs/2004MNRAS.348L..54C}
}

@ARTICLE{gaskell_2004,
       author = {{Gaskell}, C. Martin and {Goosmann}, Ren{\'e} W. and {Antonucci}, Robert R.~J. and {Whysong}, David H.},
        title = "{The Nuclear Reddening Curve for Active Galactic Nuclei and the Shape of the Infrared to X-Ray Spectral Energy Distribution}",
      journal = {\apj},
         year = 2004,
        month = nov,
       volume = {616},
       number = {1},
        pages = {147-156},
          doi = {10.1086/423885},
archivePrefix = {arXiv},
       eprint = {astro-ph/0309595},
 primaryClass = {astro-ph},
       adsurl = {https://ui.adsabs.harvard.edu/abs/2004ApJ...616..147G}
}

@ARTICLE{crenshaw_2001,
       author = {{Crenshaw}, D.~M. and {Kraemer}, S.~B. and {Bruhweiler}, F.~C. and {Ruiz}, J.~R.},
        title = "{Absorption and Reddening in the Seyfert Galaxy NGC 3227}",
      journal = {\apj},
         year = 2001,
        month = jul,
       volume = {555},
       number = {2},
        pages = {633-640},
          doi = {10.1086/321522},
archivePrefix = {arXiv},
       eprint = {astro-ph/0102404},
 primaryClass = {astro-ph},
       adsurl = {https://ui.adsabs.harvard.edu/abs/2001ApJ...555..633C}
}

@ARTICLE{crenshaw_2002,
       author = {{Crenshaw}, D.~M. and {Kraemer}, S.~B. and {Turner}, T.~J. and {Collier}, S. and {Peterson}, B.~M. and {Brandt}, W.~N. and {Clavel}, J. and {George}, I.~M. and {Horne}, K. and {Kriss}, G.~A. and {Mathur}, S. and {Netzer}, H. and {Pogge}, R.~W. and {Pounds}, K.~A. and {Romano}, P. and {Shemmer}, O. and {Wamsteker}, W.},
        title = "{Reddening, Emission-Line, and Intrinsic Absorption Properties in the Narrow-Line Seyfert 1 Galaxy Arakelian 564}",
      journal = {\apj},
         year = 2002,
        month = feb,
       volume = {566},
       number = {1},
        pages = {187-194},
          doi = {10.1086/338058},
archivePrefix = {arXiv},
       eprint = {astro-ph/0110303},
 primaryClass = {astro-ph},
       adsurl = {https://ui.adsabs.harvard.edu/abs/2002ApJ...566..187C}
}

@ARTICLE{giveon_1999,
       author = {{Giveon}, Uriel and {Maoz}, Dan and {Kaspi}, Shai and {Netzer}, Hagai and {Smith}, Paul S.},
        title = "{Long-term optical variability properties of the Palomar-Green quasars}",
      journal = {\mnras},
         year = 1999,
        month = jul,
       volume = {306},
       number = {3},
        pages = {637-654},
          doi = {10.1046/j.1365-8711.1999.02556.x},
archivePrefix = {arXiv},
       eprint = {astro-ph/9902254},
 primaryClass = {astro-ph},
       adsurl = {https://ui.adsabs.harvard.edu/abs/1999MNRAS.306..637G}
}

@ARTICLE{shapovalova_2010,
       author = {{Shapovalova}, A.~I. and {Popovi{\'c}}, L. {\v{C}}. and {Burenkov}, A.~N. and {Chavushyan}, V.~H. and {Ili{\'c}}, D. and {Kova{\v{c}}evi{\'c}}, A. and {Bochkarev}, N.~G. and {Le{\'o}n-Tavares}, J.},
        title = "{Long-term variability of the optical spectra of NGC 4151. II. Evolution of the broad H{\ensuremath{\alpha}} and H{\ensuremath{\beta}} emission-line profiles}",
      journal = {\aap},
         year = 2010,
        month = jan,
       volume = {509},
          eid = {A106},
        pages = {A106},
          doi = {10.1051/0004-6361/200912311},
archivePrefix = {arXiv},
       eprint = {0910.2980},
 primaryClass = {astro-ph.GA},
       adsurl = {https://ui.adsabs.harvard.edu/abs/2010A&A...509A.106S}
}

@ARTICLE{popovic_2011,
       author = {{Popovi{\'c}}, L. {\v{C}}. and {Shapovalova}, A.~I. and {Ili{\'c}}, D. and {Kova{\v{c}}evi{\'c}}, A. and {Kollatschny}, W. and {Burenkov}, A.~N. and {Chavushyan}, V.~H. and {Bochkarev}, N.~G. and {Le{\'o}n-Tavares}, J.},
        title = "{Spectral optical monitoring of 3C 390.3 in 1995-2007. II. Variability of the spectral line parameters}",
      journal = {\aap},
         year = 2011,
        month = apr,
       volume = {528},
          eid = {A130},
        pages = {A130},
          doi = {10.1051/0004-6361/201016317},
archivePrefix = {arXiv},
       eprint = {1101.4867},
 primaryClass = {astro-ph.CO},
       adsurl = {https://ui.adsabs.harvard.edu/abs/2011A&A...528A.130P}
}

@ARTICLE{li_2022,
       author = {{Li}, Ruancun and {Ho}, Luis C. and {Ricci}, Claudio and {Trakhtenbrot}, Benny and {Arcavi}, Iair and {Kara}, Erin and {Hiramatsu}, Daichi},
        title = "{The Host Galaxy and Rapidly Evolving Broad-line Region in the Changing-look Active Galactic Nucleus 1ES 1927+654}",
      journal = {\apj},
         year = 2022,
        month = jul,
       volume = {933},
       number = {1},
          eid = {70},
        pages = {70},
          doi = {10.3847/1538-4357/ac714a},
archivePrefix = {arXiv},
       eprint = {2208.01797},
 primaryClass = {astro-ph.GA},
       adsurl = {https://ui.adsabs.harvard.edu/abs/2022ApJ...933...70L}
}

@ARTICLE{kim_2018,
       author = {{Kim}, D. -C. and {Yoon}, Ilsang and {Evans}, A.~S.},
        title = "{Recoiling Supermassive Black Hole in Changing-look AGN Mrk 1018}",
      journal = {\apj},
         year = 2018,
        month = jul,
       volume = {861},
       number = {1},
          eid = {51},
        pages = {51},
          doi = {10.3847/1538-4357/aac77d},
archivePrefix = {arXiv},
       eprint = {1805.05251},
 primaryClass = {astro-ph.GA},
       adsurl = {https://ui.adsabs.harvard.edu/abs/2018ApJ...861...51K}
}

@ARTICLE{korista_2004,
       author = {{Korista}, Kirk T. and {Goad}, Michael R.},
        title = "{What the Optical Recombination Lines Can Tell Us about the Broad-Line Regions of Active Galactic Nuclei}",
      journal = {\apj},
         year = 2004,
        month = may,
       volume = {606},
       number = {2},
        pages = {749-762},
          doi = {10.1086/383193},
archivePrefix = {arXiv},
       eprint = {astro-ph/0402506},
 primaryClass = {astro-ph},
       adsurl = {https://ui.adsabs.harvard.edu/abs/2004ApJ...606..749K}
}

@ARTICLE{gaskell_1984,
       author = {{Gaskell}, C.~M. and {Ferland}, G.~J.},
        title = "{Theoretical hydrogen-line ratios for the narrow-line regions of active galactic nuclei}",
      journal = {\pasp},
         year = 1984,
        month = jun,
       volume = {96},
        pages = {393-397},
          doi = {10.1086/131352},
       adsurl = {https://ui.adsabs.harvard.edu/abs/1984PASP...96..393G}
}

@ARTICLE{lu_2019b,
       author = {{Lu}, Kai-Xing and {Zhao}, Yinghe and {Bai}, Jin-Ming and {Fan}, Xu-Liang},
        title = "{Reddening of the BLR and NLR in AGNs from a systematic analysis of Balmer decrement}",
      journal = {\mnras},
         year = 2019,
        month = feb,
       volume = {483},
       number = {2},
        pages = {1722-1730},
          doi = {10.1093/mnras/sty3229},
archivePrefix = {arXiv},
       eprint = {1811.11063},
 primaryClass = {astro-ph.GA},
       adsurl = {https://ui.adsabs.harvard.edu/abs/2019MNRAS.483.1722L}
}

@ARTICLE{halpern_1983,
       author = {{Halpern}, J.~P. and {Steiner}, J.~E.},
        title = "{Low ionization active galactic nuclei : X-ray or shock heated ?}",
      journal = {\apjl},
         year = 1983,
        month = jun,
       volume = {269},
        pages = {L37-L41},
          doi = {10.1086/184051},
       adsurl = {https://ui.adsabs.harvard.edu/abs/1983ApJ...269L..37H}
}

@ARTICLE{gaskell_2017,
       author = {{Gaskell}, C. Martin},
        title = "{The case for cases B and C: intrinsic hydrogen line ratios of the broad-line region of active galactic nuclei, reddenings, and accretion disc sizes}",
      journal = {\mnras},
         year = 2017,
        month = may,
       volume = {467},
       number = {1},
        pages = {226-238},
          doi = {10.1093/mnras/stx094},
archivePrefix = {arXiv},
       eprint = {1512.09291},
 primaryClass = {astro-ph.GA},
       adsurl = {https://ui.adsabs.harvard.edu/abs/2017MNRAS.467..226G}
}

@BOOK{osterbrock_2006,
       author = {{Osterbrock}, Donald E. and {Ferland}, Gary J.},
        title = "{Astrophysics of Gaseous Nebulae and Active Galactic Nuclei}",
      edition = {2nd},
         year = 2006,
    publisher = {Sausalito, CA: Univ. Science Books},
          doi = {10.1007/978-94-009-0963-2},
        adsurl = {https://ui.adsabs.harvard.edu/abs/2006agna.book.....O}
}

@ARTICLE{netzer_1975,
       author = {{Netzer}, H.},
        title = "{Physical conditions in active nuclei-I. The Balmer decrement}",
      journal = {\mnras},
         year = 1975,
        month = may,
       volume = {171},
        pages = {395-406},
          doi = {10.1093/mnras/171.2.395},
       adsurl = {https://ui.adsabs.harvard.edu/abs/1975MNRAS.171..395N}
}

@ARTICLE{davidson_1979,
       author = {{Davidson}, Kris and {Netzer}, Hagai},
        title = "{The emission lines of quasars and similar objects}",
      journal = {Reviews of Modern Physics},
         year = 1979,
        month = oct,
       volume = {51},
       number = {4},
        pages = {715-766},
          doi = {10.1103/RevModPhys.51.715},
       adsurl = {https://ui.adsabs.harvard.edu/abs/1979RvMP...51..715D}
}

@ARTICLE{dong_2008,
       author = {{Dong}, Xiaobo and {Wang}, Tinggui and {Wang}, Jianguo and {Yuan}, Weimin and {Zhou}, Hongyan and {Dai}, Haifeng and {Zhang}, Kai},
        title = "{Broad-line Balmer decrements in blue active galactic nuclei}",
      journal = {\mnras},
         year = 2008,
        month = jan,
       volume = {383},
       number = {2},
        pages = {581-592},
          doi = {10.1111/j.1365-2966.2007.12560.x},
archivePrefix = {arXiv},
       eprint = {0710.1458},
 primaryClass = {astro-ph},
       adsurl = {https://ui.adsabs.harvard.edu/abs/2008MNRAS.383..581D}
}

@ARTICLE{lamura_2007,
       author = {{La Mura}, G. and {Popovi{\'c}}, L. {\v{C}}. and {Ciroi}, S. and {Rafanelli}, P. and {Ili{\'c}}, D.},
        title = "{Detailed Analysis of Balmer Lines in a Sloan Digital Sky Survey Sample of 90 Broad-Line Active Galactic Nuclei}",
      journal = {\apj},
         year = 2007,
        month = dec,
       volume = {671},
       number = {1},
        pages = {104-117},
          doi = {10.1086/522821},
       adsurl = {https://ui.adsabs.harvard.edu/abs/2007ApJ...671..104L}
}

@ARTICLE{sriram_2022,
       author = {{Sriram}, K. and {Nour}, D. and {Choi}, C.~S.},
        title = "{Influence of Comptonization region over the ambiance of accretion disc in active galactic nucleus}",
      journal = {\mnras},
         year = 2022,
        month = mar,
       volume = {510},
       number = {3},
        pages = {3222-3235},
          doi = {10.1093/mnras/stab3610},
archivePrefix = {arXiv},
       eprint = {2112.04180},
 primaryClass = {astro-ph.HE},
       adsurl = {https://ui.adsabs.harvard.edu/abs/2022MNRAS.510.3222S}
}

@ARTICLE{brooks_2025,
       author = {{Brooks}, Madisyn and {Simons}, Raymond C. and {Trump}, Jonathan R. and {Taylor}, Anthony J. and {Bagley}, Micaela B. and {Backhaus}, Bren and {Davis}, Kelcey and {Buat}, V{\'e}ronique and {Cleri}, Nikko J. and {de la Vega}, Alexander and {Finkelstein}, Steven L. and {Hirschmann}, Michaela and {Holwerda}, Benne W. and {Kocevski}, Dale D. and {Koekemoer}, Anton M. and {Lucas}, Ray A. and {Pacucci}, Fabio and {Seill{\'e}}, Lise-Marie},
        title = "{Here There Be (Dusty) Monsters: High-redshift Active Galactic Nuclei Are Dustier than Their Hosts}",
      journal = {\apj},
         year = 2025,
        month = jun,
       volume = {986},
       number = {2},
          eid = {177},
        pages = {177},
          doi = {10.3847/1538-4357/addac4},
archivePrefix = {arXiv},
       eprint = {2410.07340},
 primaryClass = {astro-ph.GA},
       adsurl = {https://ui.adsabs.harvard.edu/abs/2025ApJ...986..177B}
}

@ARTICLE{2013A&A...558A..33A,
       author = {{Astropy Collaboration} and {Robitaille}, Thomas P. and {Tollerud}, Erik J. and {Greenfield}, Perry and {Droettboom}, Michael and {Bray}, Erik and {Aldcroft}, Tom and {Davis}, Matt and {Ginsburg}, Adam and {Price-Whelan}, Adrian M. and {Kerzendorf}, Wolfgang E. and {Conley}, Alexander and {Crighton}, Neil and {Barbary}, Kyle and {Muna}, Demitri and {Ferguson}, Henry and {Grollier}, Fr{\'e}d{\'e}ric and {Parikh}, Madhura M. and {Nair}, Prasanth H. and {Unther}, Hans M. and {Deil}, Christoph and {Woillez}, Julien and {Conseil}, Simon and {Kramer}, Roban and {Turner}, James E.~H. and {Singer}, Leo and {Fox}, Ryan and {Weaver}, Benjamin A. and {Zabalza}, Victor and {Edwards}, Zachary I. and {Azalee Bostroem}, K. and {Burke}, D.~J. and {Casey}, Andrew R. and {Crawford}, Steven M. and {Dencheva}, Nadia and {Ely}, Justin and {Jenness}, Tim and {Labrie}, Kathleen and {Lim}, Pey Lian and {Pierfederici}, Francesco and {Pontzen}, Andrew and {Ptak}, Andy and {Refsdal}, Brian and {Servillat}, Mathieu and {Streicher}, Ole},
        title = "{Astropy: A community Python package for astronomy}",
      journal = {\aap},
         year = 2013,
        month = oct,
       volume = {558},
          eid = {A33},
        pages = {A33},
          doi = {10.1051/0004-6361/201322068},
archivePrefix = {arXiv},
       eprint = {1307.6212},
 primaryClass = {astro-ph.IM},
       adsurl = {https://ui.adsabs.harvard.edu/abs/2013A&A...558A..33A}
}

@ARTICLE{2018AJ....156..123A,
       author = {{Astropy Collaboration} and {Price-Whelan}, A.~M. and {Sip{\H{o}}cz}, B.~M. and {G{\"u}nther}, H.~M. and {Lim}, P.~L. and {Crawford}, S.~M. and {Conseil}, S. and {Shupe}, D.~L. and {Craig}, M.~W. and {Dencheva}, N. and {Ginsburg}, A. and {VanderPlas}, J.~T. and {Bradley}, L.~D. and {P{\'e}rez-Su{\'a}rez}, D. and {de Val-Borro}, M. and {Aldcroft}, T.~L. and {Cruz}, K.~L. and {Robitaille}, T.~P. and {Tollerud}, E.~J. and {Ardelean}, C. and {Babej}, T. and {Bach}, Y.~P. and {Bachetti}, M. and {Bakanov}, A.~V. and {Bamford}, S.~P. and {Barentsen}, G. and {Barmby}, P. and {Baumbach}, A. and {Berry}, K.~L. and {Biscani}, F. and {Boquien}, M. and {Bostroem}, K.~A. and {Bouma}, L.~G. and {Brammer}, G.~B. and {Bray}, E.~M. and {Breytenbach}, H. and {Buddelmeijer}, H. and {Burke}, D.~J. and {Calderone}, G. and {Cano Rodr{\'\i}guez}, J.~L. and {Cara}, M. and {Cardoso}, J.~V.~M. and {Cheedella}, S. and {Copin}, Y. and {Corrales}, L. and {Crichton}, D. and {D'Avella}, D. and {Deil}, C. and {Depagne}, {\'E}. and {Dietrich}, J.~P. and {Donath}, A. and {Droettboom}, M. and {Earl}, N. and {Erben}, T. and {Fabbro}, S. and {Ferreira}, L.~A. and {Finethy}, T. and {Fox}, R.~T. and {Garrison}, L.~H. and {Gibbons}, S.~L.~J. and {Goldstein}, D.~A. and {Gommers}, R. and {Greco}, J.~P. and {Greenfield}, P. and {Groener}, A.~M. and {Grollier}, F. and {Hagen}, A. and {Hirst}, P. and {Homeier}, D. and {Horton}, A.~J. and {Hosseinzadeh}, G. and {Hu}, L. and {Hunkeler}, J.~S. and {Ivezi{\'c}}, {\v{Z}}. and {Jain}, A. and {Jenness}, T. and {Kanarek}, G. and {Kendrew}, S. and {Kern}, N.~S. and {Kerzendorf}, W.~E. and {Khvalko}, A. and {King}, J. and {Kirkby}, D. and {Kulkarni}, A.~M. and {Kumar}, A. and {Lee}, A. and {Lenz}, D. and {Littlefair}, S.~P. and {Ma}, Z. and {Macleod}, D.~M. and {Mastropietro}, M. and {McCully}, C. and {Montagnac}, S. and {Morris}, B.~M. and {Mueller}, M. and {Mumford}, S.~J. and {Muna}, D. and {Murphy}, N.~A. and {Nelson}, S. and {Nguyen}, G.~H. and {Ninan}, J.~P. and {N{\"o}the}, M. and {Ogaz}, S. and {Oh}, S. and {Parejko}, J.~K. and {Parley}, N. and {Pascual}, S. and {Patil}, R. and {Patil}, A.~A. and {Plunkett}, A.~L. and {Prochaska}, J.~X. and {Rastogi}, T. and {Reddy Janga}, V. and {Sabater}, J. and {Sakurikar}, P. and {Seifert}, M. and {Sherbert}, L.~E. and {Sherwood-Taylor}, H. and {Shih}, A.~Y. and {Sick}, J. and {Silbiger}, M.~T. and {Singanamalla}, S. and {Singer}, L.~P. and {Sladen}, P.~H. and {Sooley}, K.~A. and {Sornarajah}, S. and {Streicher}, O. and {Teuben}, P. and {Thomas}, S.~W. and {Tremblay}, G.~R. and {Turner}, J.~E.~H. and {Terr{\'o}n}, V. and {van Kerkwijk}, M.~H. and {de la Vega}, A. and {Watkins}, L.~L. and {Weaver}, B.~A. and {Whitmore}, J.~B. and {Woillez}, J. and {Zabalza}, V. and {Astropy Contributors}},
        title = "{The Astropy Project: Building an Open-science Project and Status of the v2.0 Core Package}",
      journal = {\aj},
         year = 2018,
        month = sep,
       volume = {156},
       number = {3},
          eid = {123},
        pages = {123},
          doi = {10.3847/1538-3881/aabc4f},
archivePrefix = {arXiv},
       eprint = {1801.02634},
 primaryClass = {astro-ph.IM},
       adsurl = {https://ui.adsabs.harvard.edu/abs/2018AJ....156..123A}
}

@ARTICLE{macleod_2016,
       author = {{MacLeod}, Chelsea L. and {Ross}, Nicholas P. and {Lawrence}, Andy and {Goad}, Mike and {Horne}, Keith and {Burgett}, William and {Chambers}, Ken C. and {Flewelling}, Heather and {Hodapp}, Klaus and {Kaiser}, Nick and {Magnier}, Eugene and {Wainscoat}, Richard and {Waters}, Christopher},
        title = "{A systematic search for changing-look quasars in SDSS}",
      journal = {\mnras},
         year = 2016,
        month = mar,
       volume = {457},
       number = {1},
        pages = {389-404},
          doi = {10.1093/mnras/stv2997},
archivePrefix = {arXiv},
       eprint = {1509.08393},
 primaryClass = {astro-ph.GA},
       adsurl = {https://ui.adsabs.harvard.edu/abs/2016MNRAS.457..389M}
}

@ARTICLE{sheng_2017,
       author = {{Sheng}, Zhenfeng and {Wang}, Tinggui and {Jiang}, Ning and {Yang}, Chenwei and {Yan}, Lin and {Dou}, Liming and {Peng}, Bo},
        title = "{Mid-infrared Variability of Changing-look AGNs}",
      journal = {\apjl},
         year = 2017,
        month = sep,
       volume = {846},
       number = {1},
          eid = {L7},
        pages = {L7},
          doi = {10.3847/2041-8213/aa85de},
archivePrefix = {arXiv},
       eprint = {1707.02686},
 primaryClass = {astro-ph.GA},
       adsurl = {https://ui.adsabs.harvard.edu/abs/2017ApJ...846L...7S}
}

@ARTICLE{wang_2025,
       author = {{Wang}, Shu and {Woo}, Jong-Hak and {Gallo}, Elena and {Son}, Donghoon and {Yang}, Qian and {Jin}, Junjie and {Guo}, Hengxiao and {Kong}, Minzhi},
        title = "{Dormancy and Reawakening over Years: Eight New Recurrent Changing-look AGNs}",
      journal = {\apj},
         year = 2025,
        month = mar,
       volume = {981},
       number = {2},
          eid = {129},
        pages = {129},
          doi = {10.3847/1538-4357/adadf3},
archivePrefix = {arXiv},
       eprint = {2410.15587},
 primaryClass = {astro-ph.GA},
       adsurl = {https://ui.adsabs.harvard.edu/abs/2025ApJ...981..129W}
}

@ARTICLE{baron_2016,
       author = {{Baron}, Dalya and {Stern}, Jonathan and {Poznanski}, Dovi and {Netzer}, Hagai},
        title = "{Evidence That Most Type-1 AGNs are Reddened by Dust in the Host ISM}",
      journal = {\apj},
         year = 2016,
        month = nov,
       volume = {832},
       number = {1},
          eid = {8},
        pages = {8},
          doi = {10.3847/0004-637X/832/1/8},
archivePrefix = {arXiv},
       eprint = {1603.06948},
 primaryClass = {astro-ph.GA},
       adsurl = {https://ui.adsabs.harvard.edu/abs/2016ApJ...832....8B}
}

@PHDTHESIS{halpern_1982,
       author = {{Halpern}, J.~P.},
        title = "{X-ray spectra of active galactic nuclei}",
       school = {Harvard University, Massachusetts},
         year = 1982,
        month = mar,
       adsurl = {https://ui.adsabs.harvard.edu/abs/1982PhDT.........4H}
}
\bibliographystyle{aasjournalv7}

\end{document}